\documentclass[
 reprint,
 amsmath,amssymb,
 superscriptaddress,
 nofootinbib,
]{revtex4-1}

\usepackage[all,defaultlines=3]{nowidow} % nombre max de lignes orphelines

\usepackage[section]{placeins} % float barriers
\usepackage{color} %colors
\usepackage{cancel} %bar some terms in equations
\usepackage{lineno} % adding line number
\usepackage[utf8]{inputenc} %french accents
\usepackage{natbib} %bibli
\usepackage{csquotes} %text quoting
\usepackage[caption = false]{subfig} %subfigure
\usepackage{graphicx}% Include figure files
\usepackage{dcolumn}% Align table columns on decimal point
\usepackage{bm}% bold math
\usepackage{textcomp}
\usepackage{upgreek}
\usepackage{hyperref} % liens internes
\hypersetup{colorlinks=true, linkcolor=blue, filecolor=magenta, urlcolor=blue}  % liens internes
\usepackage{float}
\usepackage{textcomp}

\usepackage[table,xcdraw]{xcolor}
\usepackage[normalem]{ulem}
\useunder{\uline}{\ul}{}

\begin{document}

\title{Thermal dissipation as both the strength and weakness of matter.\\A material failure prediction by monitoring creep}

\begin{abstract}
In any domain involving some stressed solids, that is, from seismology to general engineering, the strength of matter is a paramount feature to understand. We here discuss the ability of a simple thermally activated sub-critical model, which includes the auto-induced thermal evolution of cracks tips, to predict the catastrophic failure of a vast range of materials. It is in particular shown that the intrinsic surface energy barrier, for breaking the atomic bonds of many solids, can be easily deduced from the slow creeping dynamics of a crack. This intrinsic barrier is however higher than the macroscopic load threshold at which brittle matter brutally fails, possibly as a result of thermal activation and of a thermal weakening mechanism. We propose a novel method to compute this macroscopic energy release rate of rupture, $G_a$, solely from monitoring slow creep, and we show that this reproduces the experimental values within $50$\% accuracy over twenty different materials, and over more than four decades of fracture energy.
\end{abstract}   

\author{Tom Vincent-Dospital}
\email{vincentdospitalt@unistra.fr}
\affiliation{Université de Strasbourg, CNRS, ITES UMR 7063, Strasbourg F-67084, France}
\affiliation{SFF Porelab, The Njord Centre, Department of physics, University of Oslo, Norway}

\author{Renaud Toussaint}
\email{renaud.toussaint@unistra.fr}
\affiliation{Université de Strasbourg, CNRS, ITES UMR 7063, Strasbourg F-67084, France}
\affiliation{SFF Porelab, The Njord Centre, Department of physics, University of Oslo, Norway}

\author{Alain Cochard}
\affiliation{Université de Strasbourg, CNRS, ITES UMR 7063, Strasbourg F-67084, France}

\author{Eirik G. Flekk\o y}
\affiliation{SFF Porelab, The Njord Centre, Department of physics, University of Oslo, Norway}

\author{Knut J\o rgen M\aa l\o y}
\affiliation{SFF Porelab, The Njord Centre, Department of physics, University of Oslo, Norway}

%\date{\today} % today, but any date may be specified
 %Use showkeys class option to display
                              
\maketitle
%\tableofcontents
%\linenumbers

% ----------------------------------------------------------------------------------%

\section{Introduction:\\From slow creep to abrupt rupture}

\begin{figure}
\centering
\includegraphics[width=0.7\linewidth]{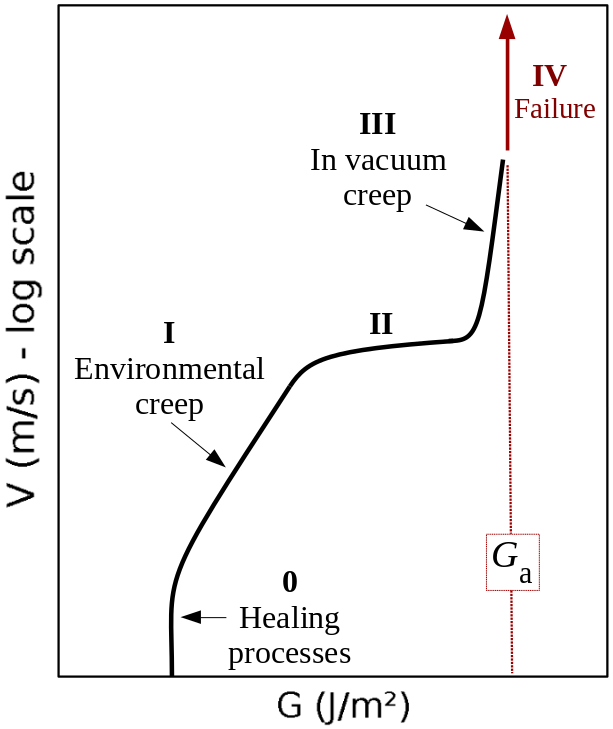}
  \caption{Summary of the different forward crack velocity regions observed in experimental  velocity curves. After \textit{Fracture of Brittle Solids}, \citet{lawn_1993}}.
 \label{BL}
\end{figure}

Although seminal, the early theoretical descriptions of crack dynamics, such as Griffith's\cite{Griffith1921} or Slepyan's\cite{Slepyan_81, Slepyan_Marder},  was somewhat binary: beyond a critical mechanical load, matter suddenly breaks. It is however acknowledged that, at load levels below the critical one, a far slower crack propagation already occurs, that will here be referred to as `creep'. This phenomenon was successfully modelled with Arrhenius-like sub-critical growth laws \cite{RICE1978, lawn_1993}, and is hence sometimes called `stress corrosion'. With the increasing number of experimental work, the description of such a slow dynamics was quickly refined, and five propagation stages were notably distinguished\cite{lawn_1993}. Let us start this manuscript by summarising them. Figure\,\ref{BL} illustrates these stages in a $V$-$G$ plot, where $V$ is the crack velocity for a given load $G$, which is the `energy release rate', that is, the energy that the fracture consumes to advance by unit surface\cite{Griffith1921}. At stage 0, while under only a mild mechanical input, cracks do not actually propagate forward. This was notably explained by the existence of some healing processes, that there efficiently compete with the failure ones\cite{RICE1978}. From this state, when the load is increased above a given threshold, some slow fracture growth starts to be observed (stage I). The propagation velocity $V$ increases exponentially with the crack's energy release rate $G$. In a sub-critical (i.e., Arrhenius-like) description, it implies that $V$ is to first order explained by an activation mechanism dependent on $G$, in a chemical-like rupture reaction\cite{kinetics}. Logically, this regime was observed to also depend on the surrounding temperature and on the fluid which is present in the fracture\cite{creepBaud}, which affects the chemical reaction involved in molecular bond breaking. When reaching a faster propagation, some velocity plateau might then hold (stage II), possibly as the transport of fluid corrosive elements toward the tip cannot efficiently cope with the crack advance. Such plateau is, in this case, only a transition to a sub-critical growth `in-vacuum-condition', where the dynamics becomes notably insensitive to the fracture fluid (stage III). Finally, when a particular threshold is reached for the energy release rate, the velocity jumps to a far quicker regime: the material fails (stage IV). We will denote\footnote{This is usually referred to as $G_c$ in experiments, since it corresponds to the value of the macroscopic energy release rate at which the velocity of fracture propagation jumps to much higher values. By contrast, in this article and our previous works, we made the choice to design as $G_c$ a microscopic property, which corresponds to an actual energy barrier in the rupture process, and we consequently used a different notation, $G_a$, for the load at which cracks avalanche to a fast phase, as a result of a boosted thermal activation (i.e.,\,see section\,\ref{section_model}). This notation choice is further discussed in section\,\ref{section_pred}.} this threshold $G_a$ in J\,m\textsuperscript{-2}, with `$a$' standing for `avalanche'.\\
In this work, we will show how studying the slow creep regime allows to predict this particular failure load. This can lead to methods to characterise natural or lowly controlled materials, where the critical energy release rate $G_a$ is not well known a priori, but where the monitoring of creep allows to infer it.
In a previous study\cite{TVD1}, we indeed proposed a unifying model of the slow creep and the fast regime, holding a precise quantification of the energy budget and the heating of the crack tip, which is coupled with an Arrhenius-type activation law. We have shown how it accounts, in some polymers \cite{TVD2}, for seven decades of propagation velocities and for the transition, at the avalanche load, from creep to sudden failure.
Here, we present how well this thermodynamics based model can predict the threshold $G_a$ for a broad range of materials, by comparing its forecasts to actual experimental failure thresholds from twenty data sets from the literature. By doing so, one can actually identify the microscopic rupture energy of the breaking bonds, $G_c$, and show how this quantity is related to, yet different from, the macroscopic $G_a$. The agreement between the predictions and the realisation is obtained for materials spanning more than 4 orders of magnitude in energy release rate, indicating the robustness of this description among different types of materials and the versatility of the theoretical framework.

\section{The thermal weakening model\label{section_model}}

\begin{figure}
\centering
\includegraphics[width=0.75\linewidth]{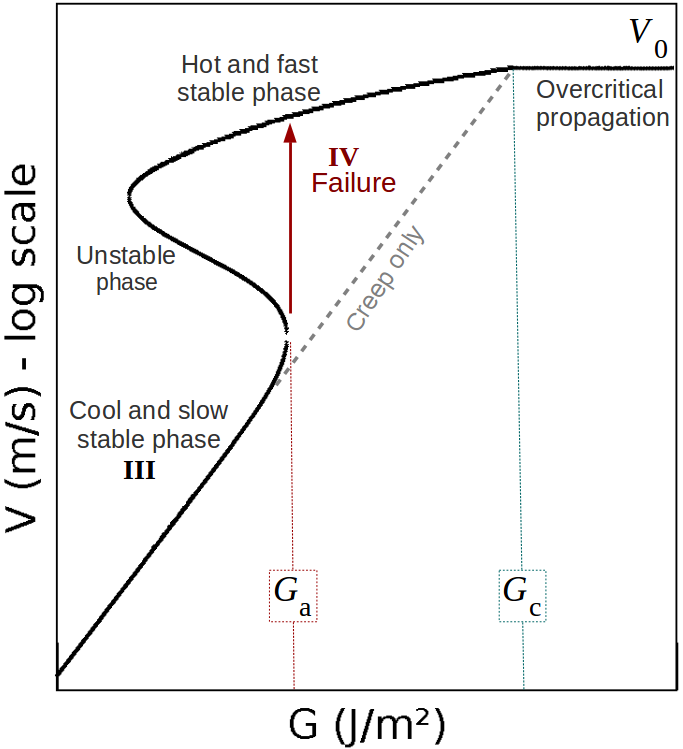}
  \caption{Modelled crack velocity as a function of energy release rate, as per Eqs.\,(\ref{model}) and (\ref{model2}). Stages III and IV correspond to the one labelled in Fig.\,\ref{BL}. As explained in the text, stages 0, I and II are here not covered. In our model, the failure occurs when the cracks becomes hot enough, that is, when $\Delta T\sim T_0$. The dashed line corresponds to a cold case $\Delta T\ll T_0$ in Eq.\,(\ref{model}).}
 \label{velmap}
\end{figure}
We consider that the propagation of a crack follows an Arrhenius sub-critical growth law, in which the temperature term accounts for the induced heat generated at the plastic crack tip\cite{orowan1954,Irwin1957}. Such a model, introduced in Refs.\,\cite{TVD1} and\,\cite{TVD2}, writes as
\begin{equation}
   V = V_0 \text{ min}\left[ \exp \left( {-\cfrac{d_0^3(G_c-G)}{2l k_B (T_0+\Delta T)}} \right), 1\right]     \label{model}
\end{equation}
\begin{equation}
   \cfrac{\partial (\Delta T)}{\partial t} = \cfrac{\lambda}{C} \nabla^2 (\Delta T) + \cfrac{\phi G V}{C \pi l^2}f, \label{model2}
\end{equation}
where the first equation describes the Arrhenius growth (i.e., the term in brackets is a probability for the thermal bath to overcome an energy barrier), and the second one in the diffusion equation governing the thermal evolution around the crack front. The heat conductivity and volumetric heat capacity of the solid matrix are respectively denoted\,$\lambda$\,and\,$C$. $V_0$ is a nominal atomic speed related to the collision frequency in the thermal bath, and should typically be comparable to the mechanical wave velocity of the studied media \cite{Slepyan_81, Freund1972}.  The activation energy is modelled proportional to $(G_c-G)$, where $G_c$ is the surface energy barrier to overcome in order to break atomic bonds. $d_0^3$ is the characteristic volume for the bonds ($d_0\sim\,1$\,\AA), $k_B$ is the Boltzmann constant, $T_0$ the ambient temperature and $\Delta T$ any variation away from it at the crack tip. A percentage $\phi$ of the power consumed per unit of crack length $GV$ is uniformly dissipated as heat over a zone of support function $f$ and of radius $l$. This heating zone is a subset of the process zone (that is, the full extent of plasticity around the tip), and we assume that it also constrains the stress level $\sigma$ at the tip, as verified in \citet{TVD2}: $\sigma\sim\sqrt{GE/l}$ (e.g., see Ref.\,\cite{lawn_1993}), where $E$ is the materials Young's modulus. This assumption is the reason why $l$ also intervenes in Eq.\,\,(\ref{model}), where the elastic energy stored in the rupturing link typically writes as $d_0^3\sigma^2/(2E)$, and is thus similar to $d_0^3G/(2l)$. In this framework, an acceleration of a fracture is thus both related to an increase of stress at the tip, which reduces the rupture energy barrier $d_0^3(G_c-G)/(2l)$, and to a related increase in volumetric internal energy ($C\Delta T$) at the fracture's head. 
\\Note that it was shown \cite{ToussaintSoft,TVD2} that at low velocities (i.e., the creep velocities we are interested in), $\Delta T$ computed from Eq.\,\,(\ref{model2}) can, more simply, be approximated to
\begin{equation}
   {\Delta T} \sim \frac{(\phi G V h) ({l/V})}{C(\pi\delta^2h)}=\frac{\phi G V}{\lambda},
   \label{velslow0}
\end{equation}
where $\phi G V h$ is the thermal power deposited along a portion of length $h$ of the crack front, $l/V$ is the characteristic time for the front heating (i.e., the time a given location of the material stays in the advancing heating zone) and $\delta\sim\sqrt{\lambda l/(\pi C V)}$ is the skin depth of heat diffusion upon the same time. 
According to Eq.\,(\ref{velslow0}), $\Delta T$ does not depend on $C$ or $l$, notably because if the crack advances slowly enough, the temperature elevation is constrained by the heat diffusion skin depth $\delta$ rather than by the size of the heat production zone, as the former is, in this case, big compared to the latter.
\\
Approximating Eqs.\,(\ref{model}) and (\ref{model2}) by their steady state solutions, two stable propagation branches are derived from this model\cite{TVD1}, as shown in Fig.\,\ref{velmap}: a fast phase, which is obtained for a hot crack tip and corresponds to the catastrophic failure of matter, and a slow one corresponding to the creep regime, when $\Delta T \ll T_0$. In between these two branches, a hysteresis situation holds with a third unstable phase. In this study, we are mainly interested in the slow to fast regime transition (i.e., that leads to quick material failure).\\
When approaching this transition, the velocity deviates from its negligible heating asymptotic expression, which is a simple exponential increase with the load $G$:
\begin{equation}
\ln \left(\frac{V}{V_0}\right) \sim (G-G_c)\left[\frac{d_0^3}{2 l k_B T_0}\right],
\label{linear}
\end{equation}
as the rise in temperature $\Delta T$ in Eq.\,(\ref{model}) becomes comparable to the room temperature $T_0$. The particular energy release rate $G_a$ is then reached, at which $\partial V/\partial G\to+\infty$, and beyond which the crack can only avalanche to a velocity which is orders of magnitude higher (see Fig.\,\ref{velmap}). Matter suddenly breaks. As a result of thermal activation, $G_a$ is actually less than the actual surface energy barrier for breaking bonds $G_c$.
\\Although rarely regarded today, such an importance of the auto-induced heat to explain brittleness was early developed\cite{Marshall_1974, Maugis1985, carbonePersson}. These studies reckon that the dissipated energy favours failure by locally softening the material at the tip (i.e., a decrease of its elastic moduli with temperature), in particular in the case of soft rubber-like materials. In the formalism of Eq.\,(\ref{model}), this view would correspond to a brutal reduction of the energy barrier $d_0^3(G_c-G)/(2l)$ at the onset of the instability, rather to (or maybe in addition to) a dramatic increase in temperature, due to chemicophysical phase changes of the matrix. Our model neglects such a softening effect and instead considers that the reaction rate for rupture is increased from the elevated temperature, only as understood by statistical physics. Of course, both views are not mutually exclusive, and one of the two mechanisms may prevail depending on the considered material, or depending on the crack velocity. Indeed, for a softening effect to be at stake, the typical time for a material temperature-related change in phase must be less than the typical warming time $l/V$ of a given location of the crack trajectory. By contrast, a change in velocity, as understood by statistical physics, should be as quick as a few atomic vibrations\cite{kinetics} ($\sim10^{13}\,$Hz) and, hence, should be more likely to explain fast cracks in hard solids. Such a qualitative discussion underlines the importance of knowing the actual rheology close to crack tips in given materials, and its sensibility to eventual temperature bursts. Note that in both approaches of thermal weakening here discussed, the $G$ value of interest (i.e., $G_a$) remains similar in its concept: the threshold for which $\Delta T$ is high enough so that a quick avalanche can be generated by a given mechanism. 
In the next sections and the rest of this manuscript, we will focus on the thermally activated, statistical, model that we have developed above, where the accurate rheology at the tip is actually considered to be of second order on the crack dynamics, and only the increase of $\Delta T$ is considered.

\section{Model predictions versus reported failures\label{section_pred}}

\begin{figure*}
\includegraphics[width=0.75\linewidth]{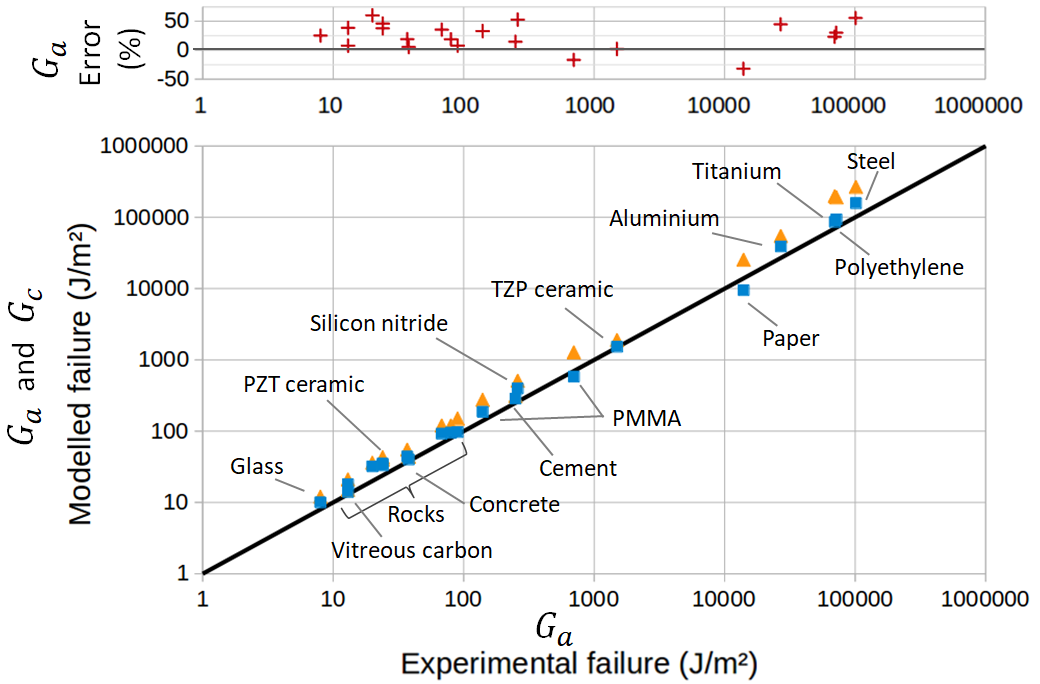}
  \centering
  \caption{(Bottom): Modelled $G_a$ thresholds (squares) and modelled surface energy barrier $G_c$ (triangles) compared to the experimental thresholds from the literature. The black line is the identity. The labels locate different materials. The unlabelled rock materials are quartz, sapphire, granite and andesite. See the ESI\textsuperscript{\dag} for an exhaustive list. (Top): Relative error on the avalanche threshold.}
 \label{crossplot}
\end{figure*}
Extensive fracturing experiments on numerous materials can be found in the literature. Hence, we can compare the model predictions of $G_a$ to some experimentally reported avalanche thresholds, that are often referred to as `critical energy release rate' or `material toughness', although it does not correspond to what is here denoted $G_c$, which is an intrinsic (microscopic) medium property not directly measurable at lab scale. In our framework, $G_c$ is called the critical energy release rate because it differentiate between an actual subcritical propagation (at $G<G_c$), where thermal activation is required for the crack to advance, and an overcrital propagation (at $G>G_c$), where enough mechanical energy is given to the crack so that thermal activation is not strictly needed. In this latter case, Eq.\,(\ref{model}) simplifies to $V=V_0$. A fast `dynamical' crack can be subcritical (see Fig.\,\ref{velmap}), in particular when $G_a<G<G_c$, and we thus refrained to call $G_a$ a critical energy release rate.\\
Note that Eq.\,(\ref{model}) does not account for all of the creep regimes summarized in Fig.\,\ref{BL}, that one can meet with an experimental test, but displays a unique low velocity slope (i.e., from Eq.\,(\ref{linear})). We have indeed discarded any healing processes, needed to explain stage 0, as they are beyond the topic of the current study. Such processes can however be included in the model \cite{TVD2}. We have also assumed no rate-limiting environmental factor, that is, no significant chemical interaction of the matrix with the fracture fluid (i.e., no stage I or II). We have hence restricted our comparison to experimental data to such a non-environmental creep case (stage III), although distinguishing it with certitude is not always straightforward. When available, we have notably preferred data sets of in-vacuum or dry (in air) experiments or with lowly corrosive (e.g., neutral pH) fracture fluids (see the creep plots in the ESI\textsuperscript{\dag} for an exhaustive list of the considered experiments, and their environmental conditions).\\
Indeed, high fluid pH or moisture tend to increase the creep dynamics of given materials (e.g.,\,\cite{quartz_creep}), presumably by altering the chemistry of the rupture process at the tip (i.e., stages I and II), and such effects are not included in our model. Note however that, when some fluid-matrix interaction does take place, the model could still be somewhat applied, if failure is preceded by a unique slope (i.e., if it occurs before the slope break between stages I and II), or after it, once clearly having entered in regime III. In the former case, the definition of the surface energy barrier $G_c$ may slightly change: from an intrinsic strength of the solid to an equivalent (lower) strength under a given chemical environment.
\\To predict $G_a$, it is of course needed to know, for each material, the values of the model constitutive parameters, that is, the parameters of Eqs.\,(\ref{model}) and (\ref{model2}), which describe the evolution of the crack velocity as a function of the applied energy release rate. 
Although they are not many, most of these parameters are not usually considered, and are hence unknown. It is however possible to estimate their order of magnitude from known material properties, or to assess them from the slow (creep) part of the loading curve. We have first considered that $V_0$ is of the order the mechanical wave velocity. It could ideally be that of the Rayleigh waves\cite{Freund1972}, but it is often simpler to instead estimate the -similar- shear wave velocity of solids, $V_S \sim \sqrt{\mu/\rho}$, as the shear modulus $\mu$ and the density $\rho$ of most materials are easily available. The heat conductivity $\lambda$ is also known in most cases, and $T_0$ is nothing but the room temperature at which a given reported experiment took place. We assume the inter-atomic space $d_0$ to be $1$\,\AA. While it could be two or three times bigger depending on the materials, which would have an order of magnitude effect on the term $d_0^3$, this uncertainty would only impact the estimation of $l$, as the ratio $d_0^3/l$ is here of importance. We indeed have to deduce $l$ and $G_c$ from the slope and intercept of the slow sub-critical growth, that is, from the two terms of Eq.\,(\ref{linear}) fitted to the experimental curves with the fit parameters $V_\text{null}$ and $b$: $\ln(V/V_\text{null})=bG$, $V_\text{null}$ being the velocity a crack would have at a null load ($G=0$), in the absence of fluid corrosion or healing effects and $b$ the slope of the creep curve of dimension m\textsuperscript{3}\,s\textsuperscript{-1}\,J\textsuperscript{-1}. This gives
%\begin{align}
\begin{equation}
   l = \frac{d_0^3}{2 b k_B T_0} \label{lin1}
\end{equation}
for the size of the thermal zone, and, for the intrinsic strength of the solid:
\begin{equation}
  G_c =  \frac{2l k_B T_0}{d_0^3} \left[\ln\left(\frac{V_0}{V_\text{null}}\right) \right].\label{lin2}
\end{equation}
This implies that we can predict $G_a$ if relying on some creep observations, that can yet be at loads far below the failure threshold. The only remaining model parameter, the percentage $\phi$ of energy converted into heat is mostly unknown. While qualitative statements, such as larger $\phi$ in metals rather than, say, polymers, are tempting, we have here arbitrarily fixed this percentage to $50$\,\% in all materials, except for a couple of instances where we could estimate it \cite{ToussaintSoft, TVD2}. For different materials, we however show how $\phi$ affects the prediction of our model in the ESI\textsuperscript{\dag}.\\
Note that, while the velocity is often reported in relation to the stress intensity factor $K$ rather than the energy release rate $G$, we have here converted from one to the other with the following relation\cite{lawn_1993}: $G \sim K^2/E$ to derive $a$ and $b$, and then $l$ and $G_c$. Backwardly, with the here proposed method, we will thus predict the toughness, $K_\text{avalanche}\sim\sqrt{E G_a}$, based on the creep measurement.\\
All the introduced parameters can now be estimated, and we did so for twenty materials for which the creeping behaviour was studied in the literature \cite{Lengline2011, TVD2, these_stephane, Yoda2001, glass_creep, saphire_creep, quartz_creep, GraniteGabbro_creep, andesite_creep, sand_creep, cement_creep, concrete_creep, zirc_tit_creep, TZP_creep, Silicon_creep, alu_creep, stainlsteel_creep, tita_creep, VitCarbon_creep}. The corresponding $G$ to $V$ curves and a table of all the inferred parameters values are shown in the ESI\textsuperscript{\dag}. We can then solve numerically the two non linear equations\,(\ref{model}) and\,(\ref{model2}), now taking into account the temperature rise $\Delta T$. In other words, for any value of the energy release rate $G$, we can compute the possible crack propagation velocities according to the model, as for instance shown in Fig.\,\ref{velmap}, where, depending on $G$, one to three velocities are possible. The inflection of the obtained curve, where $\partial V/\partial G\to+\infty$ (see Fig.\,\ref{velmap}), can be identified as $G_a$ and compared to the reported experimental thresholds. The detailed procedure is discussed and applied to two materials (PMMA and Pressure Sensitive Adhesive) in Ref.\,\cite{TVD2}. This comparison is summarised for all the media in Fig.\,\ref{crossplot}, and our model displays there a good general description of catastrophic failure. In the same figure, the surface energy barrier $G_c$ is also displayed for comparison, as well as the relative error made in the estimation of $G_a$. 

\section{Analytical approximation} \label{approx}

While, to derive the modelled $G_a$, one should compute the full crack dynamics (i.e., as displayed in Fig.\,\ref{velmap}), and search for the points where $\partial G/\partial V=0$, we explain in the ESI\textsuperscript{\dag} how Eqs.\,(\ref{model}) and (\ref{model2}) also approximately lead to
\begin{equation}
G_a \sim \frac{\lambda T_0}{\phi V_0}\frac{\exp(R_a)}{R_a},
\label{Approx2}
\end{equation}
where $R_a$ is the activation energy at the avalanche threshold counted in thermal energy units: $R_a=d_0^3(G_c-G_a)/(2l k_B T_0)$. As this ratio notably depends on $G_a$, Eq.\,(\ref{Approx2}) only implicitly defines the threshold. Although a numerical solver is there still required to compute the threshold, it is simpler and far quicker than finding the accurate solution, and potentially easy to use in potential engineering applications.\\
We show, in the ESI\textsuperscript{\dag}, how the approximation of Eq.\,(\ref{Approx2}) is a slight overestimation of the real solution, by about $0$ to $10$\%. It however gives further insight on the influence of each parameter. The remaining dependence on ($G_c-G_a$) shows that, to obtain thermal weakening, one must already be close to the actual microscopic energy barrier of rupture $G_c$. The stress at the tip thus remains an important driving mechanism of the crack propagation. The brutal failure threshold $G_a$ increases with the ambient temperature $T_0$, as a higher temperature elevation $\Delta T$ is then required at the tip to significantly overcome a higher thermal bath. Similarly, $G_a$ increases with the $\lambda/\phi$ ratio, as a high $\lambda/\phi$ indicates a rather cool crack tip, either because little energy is converted into heat (i.e., a small $\phi$) or because this energy is efficiently evacuated away from the crack (i.e., a high $\lambda$). The failure threshold also decreases with $V_0$, as, at a given load, a higher $V_0$ means a faster (and then hotter) creeping crack from a higher atomic collision frequency $V_0/d_0$.

\section{Microscopic vs macroscopic fracture energy, and local vs bulk energy dissipation\label{micro_macro}}

\begin{figure}
\includegraphics[width=1\linewidth]{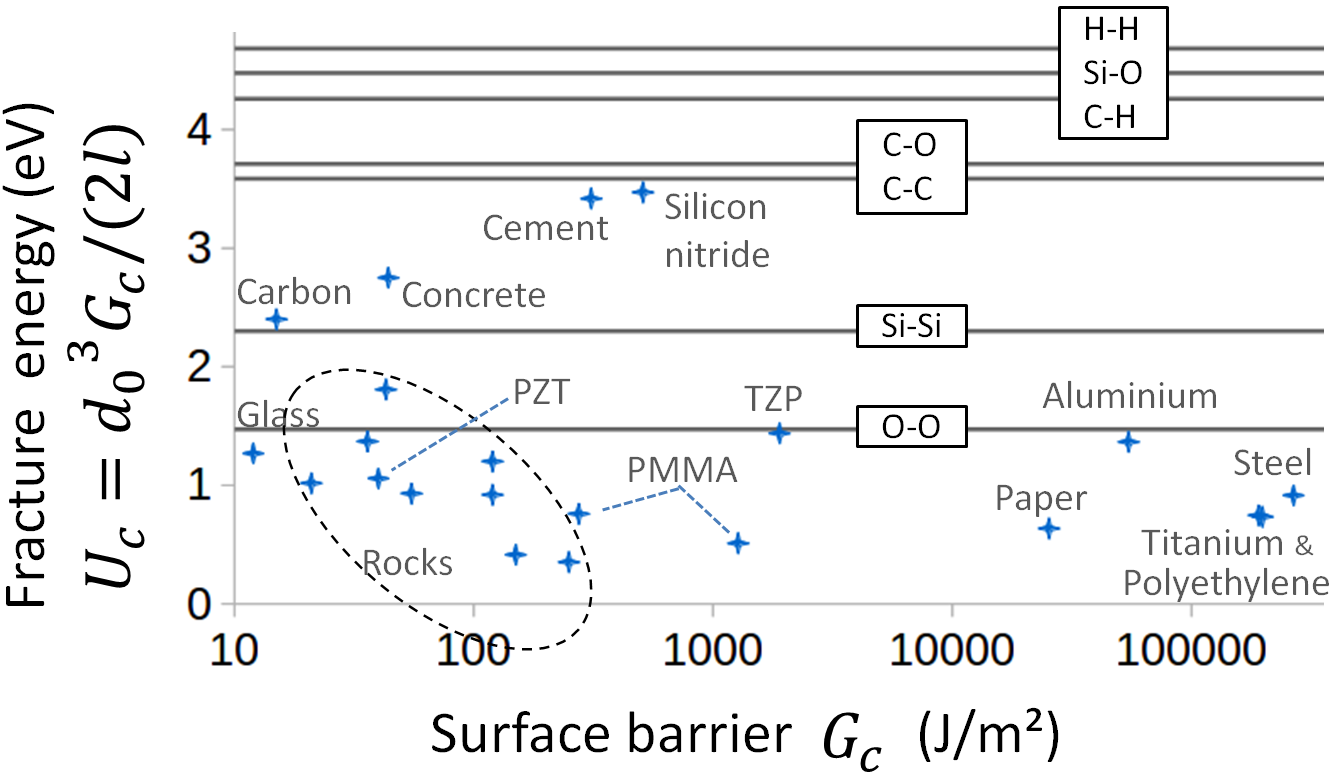}
  \caption{Microscopic fracture energy $U_c=d_0^3 G_c/(2l)$ as a function of the macroscopic energy barrier $G_c$, for the same materials as in Fig.\,\ref{crossplot}. See the ESI\textsuperscript{\dag} for the exhaustive list. Note that the accuracy of $U_c$ is not better than an order of magnitude. The horizontal lines show some typical covalent cohesion energies for comparison\cite{covalence}.}
 \label{fracener}
\end{figure}

\begin{figure}
\includegraphics[width=1\linewidth]{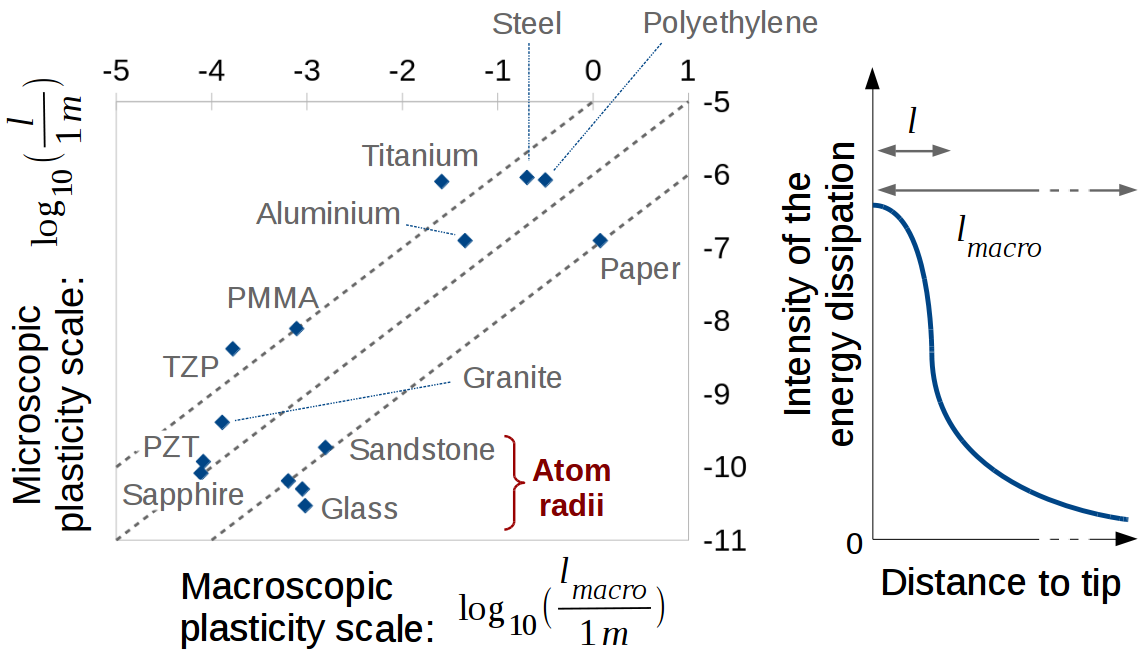}
  \caption{(Left): Core size of the process zone $l$, as understood by our model, versus macroscopic plasticity scale $l_\text{macro}$, as derived from reported tensile yield stresses. The straight lines mark a factor $10^{5}$, $10^{6}$ and $10^{7}$ between both views. The two unlabelled points in the vicinity of Glass represent Quartz and Concrete. (Right): Simplified spacial distribution of the intensity (arbitrary unit) at which energy is dissipated (i.e., plasticity) around the crack tip, as a possible explanation for the difference in scale. The graph is not to scale, as $l_\text{macro}\gg l$.}
 \label{plastsize}
\end{figure}

In Fig.\,\ref{crossplot}, one can notice that the surface energy barrier $G_c$ is always similar in order of magnitude to the rupture threshold $G_a$.
Yet, the rupture always occurs at a load less than $G_c$, with $G_a$ being about twice lower in average for all the displayed solids. We have here explained how a weakening mechanism, as the thermal view that we have here developed, allows to account for this discrepancy.\\
Having gathered various exponential creep data, and derived $l$ and $G_c$ from their slope and intercept in their $\ln{V}$ - $G$ representations, we can notably infer the intrinsic crack energy barrier in each material: $U_c=d_0^3 G_c/(2l)=k_B T_0\ln(V_0/V_\text{null})$. As shown in Fig.\,\ref{fracener}, this quantity is always in the order of  $10^{-19}$\,J$\,\sim\,1$\,eV, logically comparable to the energy level necessary to unbind single atomic covalent bounds\cite{covalence}, which confirms the relevance of a simple thermally activated model for the description of stage III (in-vacuum-like) creep. 
The actual values of $U_c$ are yet often slightly smaller than the typical covalent strength. This could derive from an averaging effect. Indeed, the Arrhenius law of Eq.\,(\ref{model}) is a mesoscopic statistical law, which we have fitted to some macroscopic measurements of crack propagation (i.e., the creep experiments). With cracks that are prone to follow the weakest paths, that possibly includes weak inter-molecular bonds (such as Van der Waals and hydrogen links) and dislocations or atomic voids (i.e., when the distance between two consecutive breaking bonds is more than a few {\aa}ngstr{\"{o}}ms), $U_c$ is likely representative of the average rupture energy in a disordered landscape. For instance, in polymers, part of the rupture shall be inter-molecular, and, in rock-type materials, the crack dynamics might benefit from the intrinsic porosity. However, due to the simplicity of the model, care should be taken when interpreting $U_c$ beyond its order of magnitude.\\
It is clear however that the value of $G_c$ varies by a factor $10^4$ for different materials, while its counterpart $U_c$ does not.
As most materials have the same $U_c$ and $d_0$, in order of magnitude, the large variability in $G_c$ (and hence in $G_a$) which is observed is, in this description, attributable entirely to the variability in the scale for the release of heat. We indeed infer that $l$ varies from the radius of a single atom, for the weakest materials, up to $1\,\upmu$m, for the ones with the highest $G_c$ (see Fig.\,\ref{plastsize}). The wider the plastic area that shields crack tips, the stronger matter is. But backwardly, we have discussed how the heat dissipation might be the root cause for dramatic ruptures in brittle solids, if the heat is not efficiently evacuated away from the rupture front. Overall, Eq.\,(\ref{model}) should be understood as:
\begin{equation}
V = V_0 \exp \left( {-\cfrac{U_c-U(\text{Load}^{[+]}\text{, Thermal radius}^{[-]}\big)}{k_B T\big(\text{Dissipation}^{[+]}\text{, Diffusion}^{[-]}\big)}} \right),
\label{model_renewed}
\end{equation}
where $U$ is the mechanical energy corresponding to the stress actually transmitted to the crack tip covalent bond of average strength $U_c$, and where [+] and [--] indicate whether the $T$ and $U$ functions are increasing or decreasing with the specified concepts (i.e., with the mechanical load, the core (thermal) dissipation radius $l$, the dissipation of heat or its diffusion). This equation emphasises that, in our formalism, thermal dissipation is both the strength of matter (from the shielding of the mechanical energy $U$ actually transmitted to the crack tip) and the weakness of matter (from the amplification in internal energy around the tip).
\\We can compare the values of $l$ with the more typical plastic radius predicted by a Dugdale view\cite{dugdale1960} of the process zone, $l_\text{macro} \sim G_c E / {\sigma_y}^2$, where $\sigma_y$ is the tensile yield stress, beyond which macroscopic samples lose their elasticity. As shown in Fig.\,\ref{plastsize}, the latter is consistently five to seven orders of magnitude higher than what we predict for $l$. This likely translates to the fact that plasticity (here understood as the dissipation of mechanical energy in any form) ought to be a rather heterogeneous phenomenon, with a greater density of energy dissipation close to the front than away from it. A higher density of energy is indeed likely to be dissipated close to the tip singularity than at the edge of the $l_\text{macro}$ plastic radius. Thus, the scale of a process zone can be characterised either by its core radius $l$, where most of the heating due to the dissipation takes place, or by its full extent $l_\text{macro}$, where the rheology becomes non elastic. While the former is to include significant thermal losses (quantified by $\phi G$), the latter can also encompass various other dissipation mechanisms for another portion of $G$, namely, the nucleation of dislocations, the release of extra heat over a greater volume around the tip, the emission of phonons and photons, or even some material change in phase (e.g., softening). In our formalism, $l$ is also the main parameter that describes the shielding of the stress near the crack tip, and this stress is indeed likely to be higher at the centre of the process zone than at its periphery, calling for $l<l_\text{macro}$. Large scale bulk energy losses (e.g., \cite{mecanochimie1, mecanochimie2}) are hence permitted by our model, and we only state that it is of second order effect on the temperature elevation and the stress level at the crack front and, hence, on the crack dynamics.\\
Besides its straight derivation from a canonical kinetics law (i.e., Eq.\,(\ref{model})), there are various indicators suggesting that, although smaller than the more common $l_\text{macro}$, the sizes $l$ that we have inverted bear a strong physical significance. For materials where $l$ is here computed to be relatively big (i.e., in the micrometer range) infrared emissions can easily be recorded around (hot) crack tips by standard resolution infrared cameras, such as in paper\cite{ToussaintSoft} or steel\cite{thermal_damage}. In the case of paper, we inferred $l$ close to the micrometric domain, which is about the dimension of the fibril forming this media. For running cracks in materials with a smaller $l$, such as PMMA or even glass, a hot temperature could also be derived by the characterisation of light emission\cite{Fuller1975, Bouchaud2012}, although its observation is not as straightforward. In glass\cite{Bouchaud2012}, in particular, the light emission lies in the visible domain, calling for blackbody\cite{blackbody} temperatures elevations of thousands of kelvins, and the light emitting zone was shown to have a nanometric radius. In PMMA, secondary microcracks tend to nucleate ahead of fast fronts from cavities/bubbles of about $100$\,nm in radius\cite{TVD2}, possibly initiated by sublimation at hot temperature of the solid matrix. In rocky materials, nanometric damages can be observed\cite{nanoparticles} on fault planes. These planes are sometimes referred to as fault mirrors, due to their glossy appearance arising from their very low roughness at the visible light wavelengths. Intense thermal effects, often referred to as `flash heating', are notably suspected for the instability of some seismic faults\cite{HeatWeak,ThermalRunaway_earthquake}.\\
Overall, some intense, extremely localised (i.e.,\,$l$), dissipation processes close to rupture fronts are thus likely to explain these various observations, and, in our framework, could also explain the brittleness of matter.

\section{Discussion and conclusion}

We have thus presented a model that gives reasonable predictions of the rupture load, over a broad range of materials. We did this with a full expression (Eqs.\,(\ref{model}) and (\ref{model2})), or in simplified forms (Eqs.\,(\ref{model}) and (\ref{velslow0}) or Eq.\,(\ref{Approx2})). This predicted load is still, however, overestimated by about $25$\% (in average for all media, see the errors in Fig.\,\ref{crossplot}).
This could derive from numerous causes. First, most of our parameters were only broadly estimated, when not arbitrarily fixed. We have in particular assumed that $G_c$ is a homogeneously distributed constant, whereas it is likely to hold some level of quenched disorder\cite{Tallakstad2011,Cochard}. In this case, the overall creep dynamics (i.e. the slow branch of Fig.\,\ref{velmap}, described by Eq.\,(\ref{linear})) would not be strongly affected, as it shall mainly depend on an average value of $G_c$. The failure, however, would be prone to occur on weaker locations \cite{TVD1}, that are controlled by a lower $G_c$, which would explain our overestimation on $G_a$. It corresponds to the common idea that the overall strength of a material is highly dependent on its heterogeneities. As discussed in section\,\ref{micro_macro}, our computation of the microscopic energy barrier $U_c$ likely represents the average of a disordered landscape in dissociation energy, as its value lies in between that of an actual covalence energy and those of weaker interactions. Additionally, we derived $G_c$ from the mean fit to experimental creep curves that, in some materials, hold some significant data dispersion (see the ESI\textsuperscript{\dag}). Some of this dispersion could arise from mesoscopic fluctuations of macroscopic fracture energy $G_c$ in these materials, and the avalanche threshold $G_a$ could be highly dependent on the standard deviation of these fluctuations. 
Another explanation for the overestimation of $G_a$ could be that the experimental error on the measurement of this parameter may in practice be important, as the avalanches occur in a regime where the crack velocity diverges with $G$, just before test samples snap at a velocity comparable to that of the mechanical waves. Hence, the last mechanical load accurately measured before rupture is, by essence, to be slightly below the actual physical threshold. Note also that, sometimes, the actual creep stage (i.e., 0 to III in Fig.\,\ref{BL}) that we fit to derive our parameters is not unambiguously identifiable on the experimental curves, while our theory does not encompass fluid-to-matrix interactions. Besides these considerations, the model is extremely simple, applying mesoscopic laws (i.e., Fourier conductivity and Arrhenius growth) at atomic scales. For instance, a propagative description \cite{non_fourier} of the heat transport (i.e., not assuming, such as Fourier diffusion, an infinite transport velocity) could be needed, due to the small time and space scales that are here considered.
Overall, a transposition of the model into a, more complicated, atomistic solver \cite{Atom3} would be beneficial.
\\Of course, the fact that the model reproduces an instability does not necessarily mean that it is the only explanation for britleness. Other models (e.g.,\,the thermal softening of the matrix around the tip\cite{carbonePersson}, or the perturbation of the stress field ahead of the crack by the rupture-induced emission of high frequency phonons\cite{Slepyan_81}) have been proposed for this instability, and, in practice, depending on the materials, more than one physical process could here be at stake. Still, the model we propose gave, in some instances \cite{TVD2}, a comprehensive explanation of the full dynamics of failure. Additionally, we have here showed how $G_c$, the intrinsic surface energy barrier of materials, shall only depend on a heat dissipation scale around the crack tip, and that the accumulation of this induced heat is effectively reducing the mechanical resistance of matter ($G_a<G_c$).\\
Countering this latter effect could be a key to design advanced strong materials, in particular as some intriguingly tough solids such as graphene \cite{Graphene_creep, Graphene_cond} or arachnid silks \cite{Spider_Cond}, are indeed very conductive. Interestingly, the conductivity of spider threads even increases with deformation \cite{Spider_Cond}, which could be a nature made adaptive defence mechanism for the stability of nets, whenever they are pressurized. Replicating such a behaviour with a man-made material would then be an important achievement that could lead to high performance cables or bulk materials. For instance, a first step could be the engineering of highly conductive atomic networks, integrated into strong solid matrices, thus limiting any local rise in temperature that could weaken matter.
\\A more down-to-earth application of the model could be the monitoring of structures and infrastructures, as we have shown how their creep rate can be used to predict their failure. This would be of particular interest for bodies that have aged in uncontrolled conditions, in which the change in mechanical properties becomes uncertain with time, but could be inverted from their creep.
\\Finally, and although we have only treated about fracture in mode I, we suggest that most of the effects that we have discussed shall be valid for mixed-mode fracturing and solid friction. The latter is also suspected to hold some non negligible, thermal related, weakening mechanisms \cite{HeatWeak,ThermalRunaway_earthquake}, which could notably be a key in geophysics and in understanding the stability of seismic faults. In particular, when increasing the background temperature $T_0$, it was shown that the model holds a critical point, beyond which not enough heat can be generated to trigger instabilities in the dynamics of cracks\cite{TVD1}, which may physically explain the brittle-ductile transition in the Earth's crust\cite{Scholz1988,Aharonov_Scholz}, below which rocks tend to flow rather than break.\\

%\newpage
\section*{Acknowledgements}

\noindent The authors declare no competing financial interests in the publishing of this work and acknowledge the support of the IRP France-Norway D-FFRACT, of the Universities of Strasbourg and Oslo and of the CNRS INSU ALEAS program. We thank the Research Council of Norway through its Centres of Excellence funding scheme, project number 262644.

% ----------------------------------------------------------------------------------%

\FloatBarrier
\bibliographystyle{unsrtnat}
\bibliography{scr.bib}

% ----------------------------------------------------------------------------------%

\end{document}

%\title{A prediction for the hot break through matter}
%\title{Is breaking through matter a hot matter?\\A material failure prediction by monitoring creep\\-----------------------------------------------------------------------\\Electronic Supplementary Information (ESI)}
\title{Thermal dissipation as both the strength and weakness of matter.\\A material failure prediction by monitoring creep\\-----------------------------------------------------------------------\\Electronic Supplementary Information (ESI)}

%\begin{abstract}
%\end{abstract}   

\author{Tom Vincent-Dospital}
\email{tom.vincent-dospital@fys.uio.no}
\affiliation{Université de Strasbourg, CNRS, ITES UMR 7063, Strasbourg F-67084, France}
\affiliation{SFF Porelab, The Njord Centre, Department of physics, University of Oslo, Norway}

\author{Renaud Toussaint}
\email{renaud.toussaint@unistra.fr}
\affiliation{Université de Strasbourg, CNRS, ITES UMR 7063, Strasbourg F-67084, France}
\affiliation{SFF Porelab, The Njord Centre, Department of physics, University of Oslo, Norway}

\author{Alain Cochard}
\affiliation{Université de Strasbourg, CNRS, ITES UMR 7063, Strasbourg F-67084, France}

\author{Eirik G. Flekk\o y}
\affiliation{SFF Porelab, The Njord Centre, Department of physics, University of Oslo, Norway}

\author{Knut J\o rgen M\aa l\o y}
\affiliation{SFF Porelab, The Njord Centre, Department of physics, University of Oslo, Norway}

%\date{\today} % today, but any date may be specified
\keywords{Rupture dynamics, thermal weakening, statistical physics} %Use showkeys class option to display
                              
\maketitle
\tableofcontents
%\linenumbers

% ----------------------------------------------------------------------------------%

\section{Analytical approximation\\of the avalanche threshold}

\begin{figure}
  \includegraphics[width=0.9\linewidth]{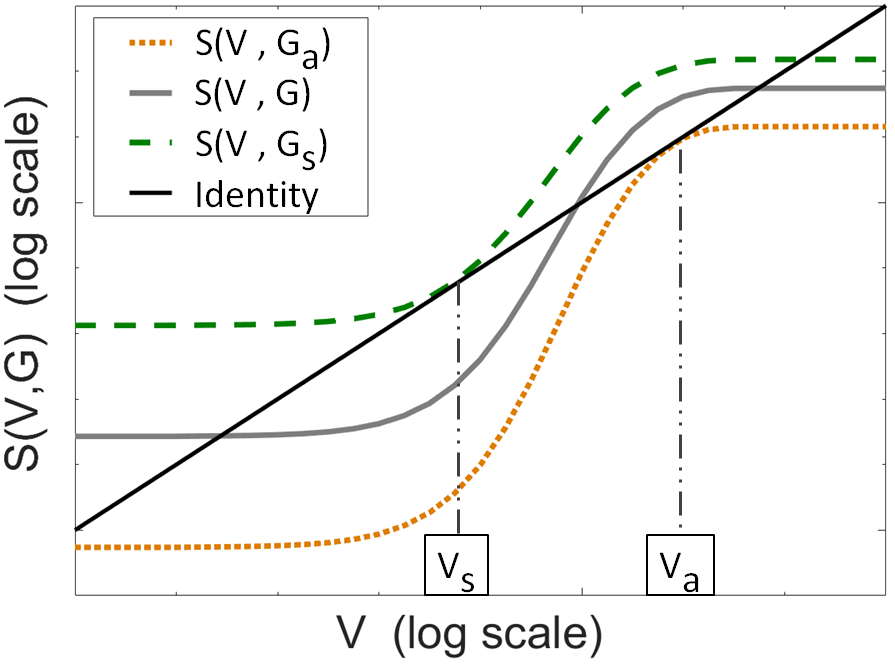}
  \caption{Representation of $V=S(V,G)$ for three values of $G$: $G_s$, $G_a$ and a mid-value between $G_s$ and $G_a$ (plain plot). The intersections of $S_G$ with the identity plot (straight line) give the possible crack velocities for a given energy release rate, as per Eq.\,(\ref{model}). The axes are not annotated for the sake of generality. See Ref.\,\cite{TVD1} for further information.}
  \label{That_is_an_S}
\end{figure}

Let us start this additional material with the analytical approximations of the temperature at a running crack tip. Assuming a quasi-constant velocity and energy release rate, some simplified expressions can indeed be derived \cite{ToussaintSoft} for $\Delta T$. At low velocity, the typical diffusion skin depth is large compared to the radius of the heat production zone ($\sqrt{\lambda l/(V \pi C)}/l \gg 1$) and the heat diffusion is hence the ruling process:
\begin{equation}
   {\Delta T}_{\text{slow}} \sim \phi G\frac{V}{\lambda}.
   \label{velslow}
\end{equation}
At high velocity, however, the rise in temperature is limited by the scale over which heat is produced and:
\begin{equation}
   {\Delta T}_{\text{fast}} \sim \frac{\phi G}{\pi C l}.
   \label{velfast}
\end{equation}
Between these two cases, and typically for $V \sim \lambda/(\pi C l)$, an intermediate regime holds:
\begin{equation}
  {\Delta T}_{\text{mid}} \sim \phi G \sqrt{\frac{V}{4\pi C\lambda l}}.
  \label{velmid}
\end{equation}
We invite the reader to a more in-depth derivation of these equations in \citet{ToussaintSoft} or \citet{TVD1}.
\\
Now that some straightforward expressions for $\Delta T$ are known, we can move on to infer $G_a$. Our model, the Arrhenius law as considered in the main manuscript, defines a function $S(V,G)$ such that $S(V,G)=V$:
\begin{equation}
   S(V,G) = V_0 \text{ min}\left[ \exp \left( {-\cfrac{\alpha^2[G_c-G]}{k_B [T_0+\Delta T(V,G)]}} \right), 1\right].
   \label{model}
\end{equation}
\\To lighten the equations that will follow, we have here denoted $\alpha^2$ the ratio $d_0^3/(2l)$. We have discussed, in the main manuscript, how this relation might have one to three solutions depending on $G$ (see Fig.\,\ref{That_is_an_S}). Two particular energy release rates mark the passages from a singular to multiple solutions: the avalanche threshold $G_a$, of interest in this study, and another threshold, $G_s$, which is the load at which an avalanche has to stop.
%(i.e., the second point where $\partial G/\partial V=0$ in Fig. \ref{velmap})
\begin{figure}
\includegraphics[width=0.9\linewidth]{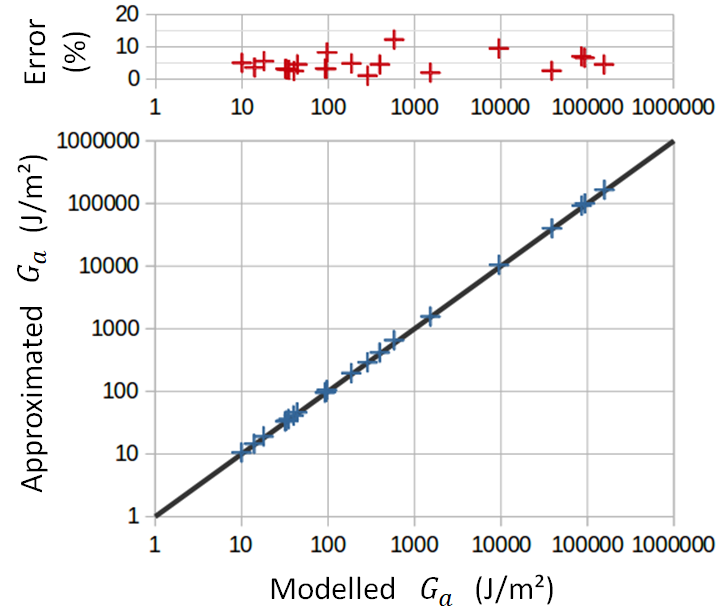}
  \caption{(Bottom): $G_a$ threshold, as approximated by Eq.\,(\ref{Approx2_repeat}) versus the accurate numerical solution of the model. The black line is the identity. (Top): Relative error from the approximation}
 \label{approxComp}
\end{figure}
\\All functions being continuously smooth, the switch from one solution to three solutions implies that $S(V,G)$ is tangent to the identity function for these two particular $G$, as illustrated in Fig \ref{That_is_an_S}.
$G_a$ and the corresponding velocity $V_a$ must therefore verify the following system of equations:\\
\begin{align}
S(V,G)=V \label{system1} \\
\frac{\partial S}{\partial V}(V,G)=1. \label{system2}
\end{align}
\\To solve this system, we assume that the transition towards the fast phase happens in a regime where the temperature elevation still increases linearly with the crack velocity (i.e., $\Delta T = {\Delta T}_\text{slow}(V,G)$ (\ref{velslow})). Equation (\ref{system2}) then becomes:\\
\begin{equation}
   \frac{\phi G \lambda \alpha^2(G_c-G)}{k_b(\lambda T_0+\phi G V)^2}S(V,G)=1.
   \label{sysresol1}
\end{equation}
\\Inserting Eq.\,(\ref{system1}) back into (\ref{sysresol1}) leads to the following quadratic equation in $V$:\\
\begin{equation}
   \left(\frac{\phi G V}{\lambda T_0}\right)^2 + \left[ 2+\frac{\alpha^2(G-G_c)}{k_b T_0}\right] \frac{\phi G V}{\lambda T_0} + 1 = 0.
   \label{sysresol2}
\end{equation}
\\While it might of course hold two solutions, only the lower one is of interest to derive the avalanche threshold $G_a$. The upper solution would indeed correspond to the `arrest' of the crack avalanche, but the initial hypothesis of $\Delta T = {\Delta T}_\text{slow}$ would there be wrong anyway, as this `arrest' occurs while on the quick (hot) propagation branch. Focusing therefore on the lower solution of (\ref{sysresol2}), we have:
\begin{equation}
V_a=\frac{T_0\lambda}{2\phi G_a}(R_a-2-R_a\sqrt{1-4/R_a}), \label{sys_aval1}
\end{equation}
\\with $R_a=\alpha^2(G_c-G_a)/(k_B T_0)$. This equation indicates at which slow velocity a crack avalanches, given the corresponding $G_a$ threshold. Substituting (\ref{sys_aval1}) in (\ref{system1}), one finally derives the equality that defines the avalanche threshold:\\
\begin{equation}
G_a \sim \frac{\lambda T_0}{2\phi V_0}\frac{R_a-2-R_a\sqrt{1-4/R_a}}{\exp \left(-2 \big/ \big[1-\sqrt{1-4/R_a}\big]\right)}.
\label{sys_aval2}
\end{equation}
\\Such an expression gives a fairly good approximation of $G_a$ as predicted by the model. The only hypothesis was indeed the validity of Eq.\,(\ref{velslow}), that is $\sqrt{\lambda l/(V_a \pi C)}/l \gg 1$ and, for the materials that we have studied in our manuscript, this ratio ranges from $300$ to $1500$. While Eq.\,(\ref{sys_aval2}) is easy to solve for $G_a$ with any numerical method, it can however be further simplified by grossly assuming that $R_a \gg 4$ and by developing the term $\sqrt{1-4/R_a}$. We thus obtain the equation presented in the manuscript:\\
\begin{equation}
G_a \sim \frac{\lambda T_0}{\phi V_0}\frac{\exp(R_a)}{R_a}.
\label{Approx2_repeat}
\end{equation}
Figure \ref{approxComp} shows the quality of the approximation for $G_a$, off by a few percents as, as shown in Tab. \ref{table}, the $R_a \gg 4$ hypothesis is not strictly valid.

\section{The arrest threshold\\(for completeness)}

Similarly, one can solve (\ref{system1}) and (\ref{system2}) at the `arrest' point: the transition from a quick regime back to the low velocity phase, occurring at the particular load $G_s$. While $G_a$ is vastly reported for a lot of materials, making it the topic of this manuscript, $G_s$ is more rarely reported, so that the following computation is given for completeness. We here assume that the transition arises when the crack cools down from the plateau temperature $\Delta T={\Delta T}_\text{fast}(G)$ (\ref{velfast}), along the intermediate slope defined by $\partial \Delta T / \partial V = \partial {\Delta T}_\text{mid}(V,G)) / \partial V$ (\ref{velmid}). We thus turn the system into a quadratic equation of $\sqrt{V}$:\\
\begin{equation}
  \left(\frac{\phi G \sqrt{V}}{4\pi \lambda Cl T_0}\right)^2 + \left(2+\frac{\alpha^2(G-G_c)}{2 k_b T_0}\right) \left(\frac{\phi G \sqrt{V}}{4\pi \lambda Cl T_0}\right) + 1 = 0,
  \label{sysresol3}
\end{equation}
\\the upper solution of which, together with Eq.\,(\ref{system1}), leads to:\\
\begin{equation}
V_s=\frac{\pi\lambda Cl{T_0}^2}{4(\phi G_s)^2}\left[R_s-4+R_s\sqrt{1-8/R_s}\right]^2 \label{sys_stop1},
\end{equation}
\\where $R_s=\alpha^2(G_c-G_s)/(k_B T_0)$. When inserting (\ref{sys_stop1}) back into (\ref{system1}), one gets:\\
\begin{equation}
\frac{4(\phi G_s)^2 V_0}{\pi\lambda Cl{T_0}^2} = \frac{\left[R_s-4+R_s\sqrt{1-8/R_s}\right]^2}{\exp\left(\cfrac{\alpha^2(G_s-G_c)}{k_b [T_0+\phi G_s /(\pi Cl)]}\right)}. \label{sys_stop2}
\end{equation}
\\Assuming that $R_s \gg 8$ and $\Delta T_\text{fast} \gg T_0$, Eq.\,(\ref{sys_stop2}) further simplifies to:
\begin{equation}
G_s \sim \frac{T_0}{\phi}\sqrt{\frac{\pi \lambda Cl}{V_0}}\exp\left(\frac{\pi ClT_0}{2\phi G_s}\right)R_s,
\end{equation}
which gives a relatively simple expression to invert for $G_s$.

%\vspace{4cm}

\section{Sensitivity of the $\phi$ parameter}

\begin{figure}[b]
\includegraphics[width=1\linewidth]{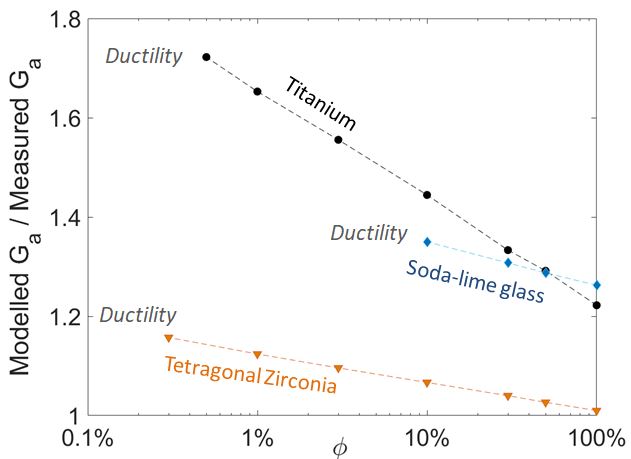}
  \caption{Variation of the model predicted failure threshold $G_a$ as a function of the heat efficiency $\phi$, relative to the experimentally reported $G_a$, for a Titanium alloy ($G_c\sim190\,$kJ\,m\textsuperscript{-2}), TZP ceramic ($G_c\sim1900\,$J\,m\textsuperscript{-2}), and soda-lime glass ($G_c\sim12\,$J\,m\textsuperscript{-2}). The considered model parameters for these materials are shown in table\,\ref{table}.}
 \label{phi}
\end{figure}
The percentage $\phi$ of energy that is converted into heat in the heat zone of radius $l$ is a parameter that we have broadly assumed to be $0.5$. In practice, this parameter is unknown, and shall likely be material dependent. In Fig.\,\ref{phi}, we show the variation in our model prediction of the failure threshold $G_a$ when varying $\phi$, for three materials spanning the whole range of considered toughnesses. The higher $\phi$, the less is $G_a$ (e.g.,\,see Eq.(\ref{Approx2_repeat})), as it makes the crack tip hotter at a lower load. As shown in the figure, when considering small values of this heat efficiency, one can also obtain an all ductile behavior\,\cite{TVD1}, as the crack becomes too cold to hold an instability in its dynamics. In this case, the modelled crack velocity only increases exponentially to $V_0$, as a function of the energy release rate $G$, and no velocity jump (defining $G_a$) is obtained.

\section{Materials creep crossplots\\and parameters table}

\begin{table*}
\begin{tabular}{l|c|c|c|c|c|c|c|c|c|}
\cline{2-10}
                                                    & \multicolumn{1}{l|}{$\lambda$ (SI)} & \multicolumn{1}{l|}{$\phi$ (-)} & \multicolumn{1}{l|}{$V_0$ (m/s)} & \multicolumn{1}{l|}{$l$ (\AA)} & \multicolumn{1}{l|}{$T_0$ (K)}  & \multicolumn{1}{l|}{$G_c$ (J/m$^2$)} & \multicolumn{1}{l|}{$G_a$ real (J/m$^2$)} & \multicolumn{1}{l|}{$G_a$ model (J/m$^2$)} & \multicolumn{1}{l|}{Ra (-)} \\ \hline
\multicolumn{1}{|l|}{Acrylic adhesive}              & 0.4                              & 1                            & 30                            & 10                        & 296                          & \textbf{150}                   & \textbf{90}                        & \textbf{97}                       & 5.7                         \\ \hline
\multicolumn{1}{|l|}{Paper}                         & 0.035                            & 0.12                         & 1300                          & 1000                        & 296                          & \textbf{25000}                 & \textbf{14000}                     & \textbf{9500}                     & 15.6                        \\ \hline
\multicolumn{1}{|l|}{Bulk PMMA}                     & 0.18                             & 0.2                          & 880                           & 80                        & 296                          & \textbf{1300}                  & \textbf{700}                       & \textbf{580}                      & 10.9                        \\ \hline
\multicolumn{1}{|l|}{Interfacial PMMA}              & 0.18                             & 0.2                          & 880                           & 8                        & 298                          & \textbf{275}                   & \textbf{140}                       & \textbf{190}                      & 13.5                        \\ \hline
\multicolumn{1}{|l|}{HD Polyethylene}                  & 0.4                              & 0.5                          & 900                           & 8500                        & 293                          & \textbf{200000}                & \textbf{70000}                     & \textbf{87000}                    & 16.6                        \\ \hline
\multicolumn{1}{|l|}{Soda lime glass}               & \cellcolor[HTML]{ECF4FF}1        & 0.5                          & 3400                          & 0.3                        & 296                          & \textbf{12}                    & \textbf{8}                       & \textbf{10}                      & 8.3                         \\ \hline
\multicolumn{1}{|l|}{Sapphire}                      & 24                               & 0.5                          & 6000                          & 0.8                        & 296                          & \textbf{36}                    & \textbf{20}                        & \textbf{32}                       & 6                         \\ \hline
\multicolumn{1}{|l|}{Quartz}                        & \cellcolor[HTML]{ECF4FF}8        & 0.5                          & 3400                          & 0.6                        & 293                          & \textbf{21}                    & \textbf{13}                        & \textbf{18}                       & 5.7                         \\ \hline
\multicolumn{1}{|l|}{Westerly Granite (ambient)}             & \cellcolor[HTML]{ECF4FF}2        & 0.5                          & 3000                          & 4                        & 293                          & \textbf{120}                   & \textbf{68}                        & \textbf{92}                      & 8.5                         \\ \hline
\multicolumn{1}{|l|}{Westerly Granite (hot)}                 & \cellcolor[HTML]{ECF4FF}2        & 0.5                          & 3000                          & 0.7                        & \cellcolor[HTML]{FFCCC9}573  & \textbf{43}                    & \textbf{24}                        & \textbf{35}                       & 6.8                         \\ \hline
\multicolumn{1}{|l|}{Kumamoto Andesite}                      & \cellcolor[HTML]{ECF4FF}1        & 0.5                          & 2200                          & 3                        & \cellcolor[HTML]{FFCCC9}330  & \textbf{120}                   & \textbf{80}                        & \textbf{97}                       & 8.8                         \\ \hline
\multicolumn{1}{|l|}{Scioto Sandstone}                     & \cellcolor[HTML]{ECF4FF}2        & 0.5                          & 2000                          & 2                        & 296                          & \textbf{55}                    & \textbf{37}                        & \textbf{44}                       & 7.3                         \\ \hline
\multicolumn{1}{|l|}{Cement paste}                        & \cellcolor[HTML]{ECF4FF}1        & 0.5                          & 2200                          & 3                        & 298                          & \textbf{310}                   & \textbf{250}                       & \textbf{280}                      & 10.7                        \\ \hline
\multicolumn{1}{|l|}{HSULP Concrete}                      & 0.8                              & 0.5                          & 3000                          & 1                        & 293                          & \textbf{44}                    & \textbf{38}                        & \textbf{40}                       & 9.9                         \\ \hline
\multicolumn{1}{|l|}{Vitreous carbon}               & \cellcolor[HTML]{ECF4FF}5        & 0.5                          & 2600                          & 0.2                        & 296                          & \textbf{15}                    & \textbf{13}                        & \textbf{14}                       & 7.2                         \\ \hline
\multicolumn{1}{|l|}{Lead Zirconate Titanate (PZT)} & \cellcolor[HTML]{ECF4FF}1        & 0.5                          & 2000                          & 1                        & 296                          & \textbf{40}                    & \textbf{24}                        & \textbf{33}                       & 11.3                        \\ \hline
\multicolumn{1}{|l|}{Tetragonal zirconia (TZP)}     & \cellcolor[HTML]{ECF4FF}2        & 0.5                          & 1600                          & 40                        & 298                          & \textbf{1900}                  & \textbf{1500}                      & \textbf{1530}                     & 10.9                         \\ \hline
\multicolumn{1}{|l|}{Silicon nitride}               & \cellcolor[HTML]{DAE8FC}30       & 0.5                          & 5500                          & 45                        & \cellcolor[HTML]{FFA19F}1573 & \textbf{510}                   & \textbf{260}                       & \textbf{400}                      & 8.9                         \\ \hline
\multicolumn{1}{|l|}{2650 T6 Aluminium alloy}               & \cellcolor[HTML]{DAE8FC}150      & 0.5                          & 3100                          & 1000                        & \cellcolor[HTML]{FFCCC9}448  & \textbf{54500}                 & \textbf{27000}                     & \textbf{39000}                    & 10.1                        \\ \hline
\multicolumn{1}{|l|}{AISI 310S Stainless Steel alloy}         & \cellcolor[HTML]{ECF4FF}14       & 0.5                          & 3000                          & 9000                        & 298                          & \textbf{265000}                & \textbf{102000}                    & \textbf{158000}                   & 13.4                        \\ \hline
\multicolumn{1}{|l|}{Ti-6A1-6V-2Sn Titanium}           & \cellcolor[HTML]{ECF4FF}7        & 0.5                          & 3100                          & 8000                        & 298                          & \textbf{190000}                & \textbf{72000}                     & \textbf{93000}                    & 14.9                        \\ \hline
\end{tabular}
\caption{Model parameters for various materials of the literature. The real and modelled $G_a$ thresholds are compared in the two former last columns. The cells colour help to highlight standing out values for $\lambda$ and $T_0$.}\label{table}
\end{table*}

\begin{figure}[H]
\includegraphics[width=0.9\linewidth]{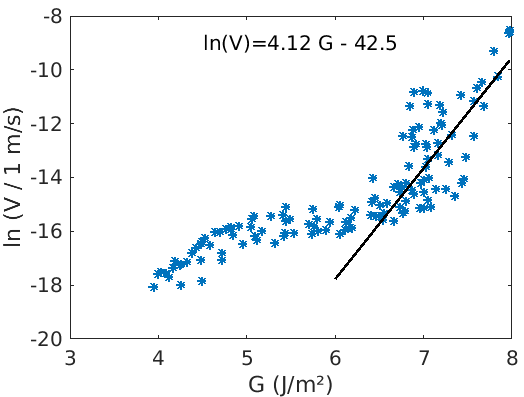}
  \caption{Creep data of dry soda-lime glass, from \citet{glass_creep}, figure 3. A rather complex creep law holds there so that we only roughly fitted the last part (i.e., stage III).}
 \label{glass_creep}
\end{figure}

\begin{figure}[H]
\includegraphics[width=0.9\linewidth]{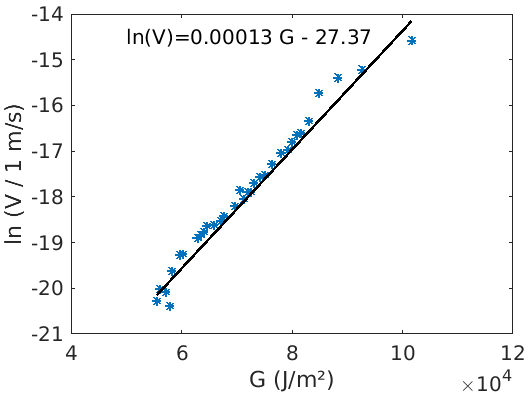}
  \caption{Creep data of AISI 310S austenitic stainless steel in air, from \citet{stainlsteel_creep}, figure 1 (uncharged plot).}
 \label{steel_creep}
\end{figure}

A summary of the model parameters considered for each media is also provided in Tab. \ref{table}. These parameters are deduced, as explained in the main manuscript, from the $V$ to $G$ creep data of these materials, shown in Fig \ref{glass_creep} to Fig \ref{steel_creep}. One can notably notice the variability in fit quality for these datasets, that of course impacts our inversion work, but also how it is not always straightforward to know to which subcritical phase the data correspond (i.e., phase I to III, from environmental induced corrosion to void-like conditions).
\begin{figure}
\includegraphics[width=0.9\linewidth]{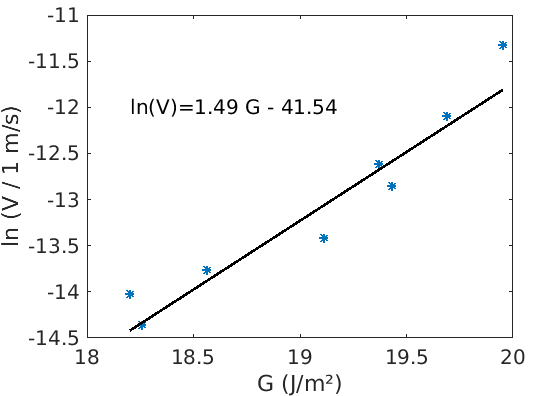}
  \caption{Creep data of dry sapphire (r-plane), from \citet{saphire_creep}.}
 \label{saphire_creep}
\end{figure}
\begin{figure}[b]
\includegraphics[width=0.9\linewidth]{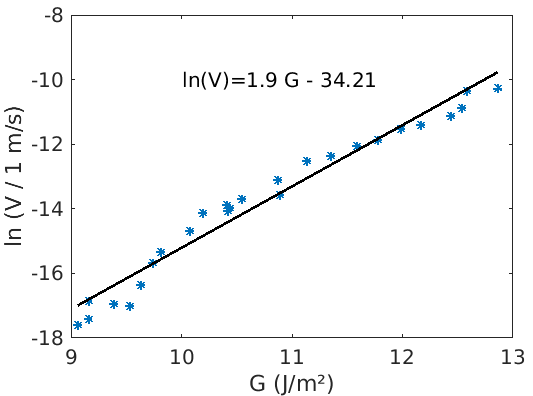}
  \caption{Creep data of quartz in vacuum, from \citet{quartz_creep}, figure 4.}
 \label{quartz_creep}
\end{figure}
\begin{figure}[b]
\includegraphics[width=0.9\linewidth]{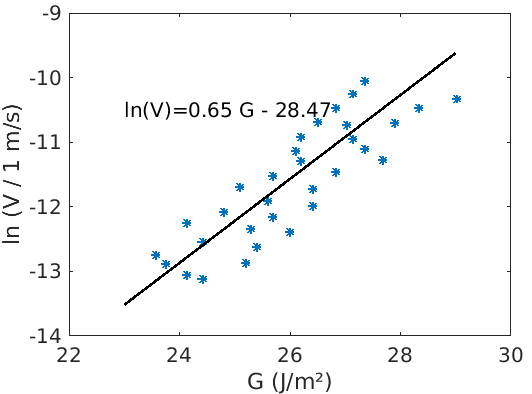}
  \caption{Creep data of Scioto sandstone, from \citet{sand_creep}, figure 3.}
 \label{sand_creep}
\end{figure}
\begin{figure}[H]
\includegraphics[width=0.9\linewidth]{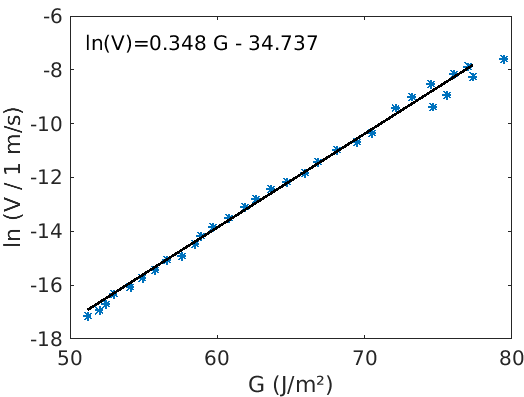}
  \caption{Creep data of Kumamoto andesite in moist air at $67$\,\textdegree\,C, from \citet{andesite_creep}, figure 9.}
 \label{andesite_creep}
\end{figure}
\begin{figure}[H]
\includegraphics[width=0.9\linewidth]{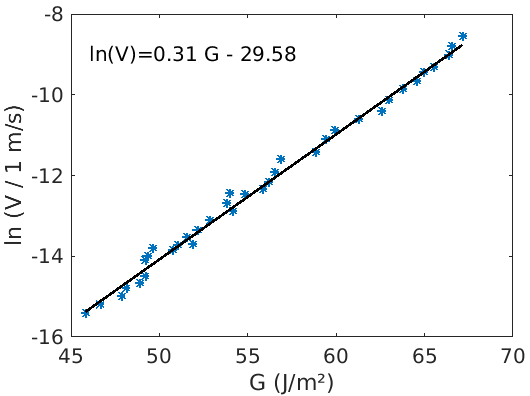}
  \caption{Creep data of Westerly granite in moist air at $20$\,\textdegree\,C, from \citet{GraniteGabbro_creep}, figure 7.}
 \label{granite_creep}
\end{figure}
\begin{figure}[H]
\includegraphics[width=0.9\linewidth]{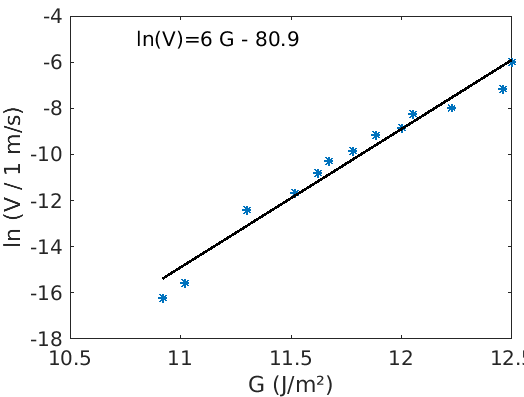}
  \caption{Creep data of vitreous carbon, from \citet{VitCarbon_creep}, figure 4.}
 \label{carbon_creep}
\end{figure}
\begin{figure}[H]
\includegraphics[width=0.9\linewidth]{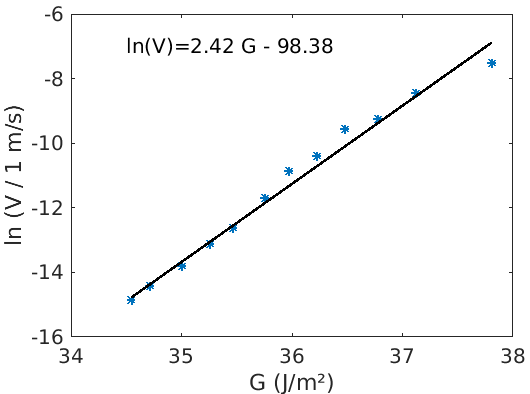}
  \caption{Creep data of high strength ultra low porosity concrete in moist air, from \citet{concrete_creep}, figure 9.}
 \label{concrete_creep}
\end{figure}
\begin{figure}[H]
\includegraphics[width=0.9\linewidth]{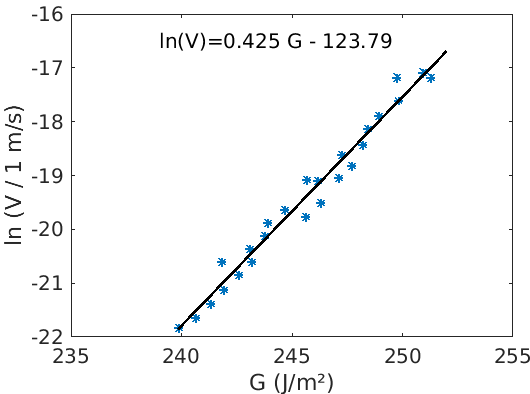}
  \caption{Creep data cement in water, from \citet{cement_creep}, figure 4a.}
 \label{cement_creep}
\end{figure}
\begin{figure}[H]
\includegraphics[width=0.9\linewidth]{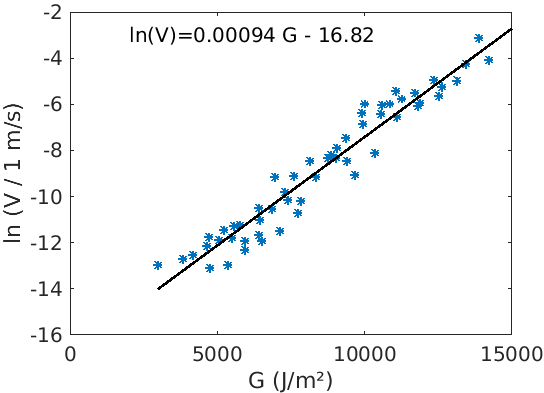}
  \caption{Creep data of paper in air, from \citet{these_stephane}, figure 3.32.}
 \label{paper_creep}
\end{figure}
\begin{figure}[H]
\includegraphics[width=0.9\linewidth]{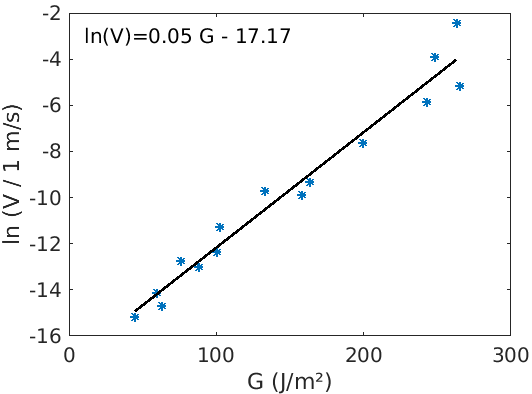}
  \caption{Creep data of hot silicon nitride at $1200$\,\textdegree\,C, from \citet{Silicon_creep}, figure 5.}
 \label{Silicon_creep}
\end{figure}
\begin{figure}[H]
\includegraphics[width=0.9\linewidth]{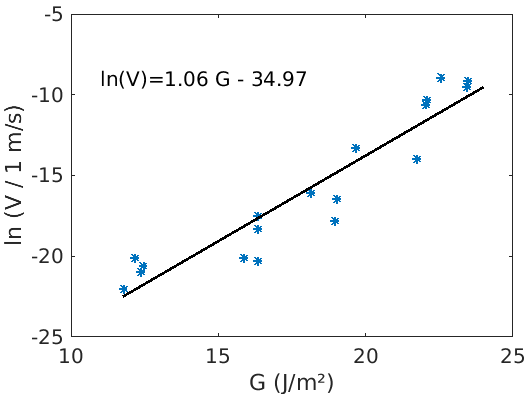}
  \caption{Creep data of Lead Zirconate Titanate at ambient conditions, from \citet{zirc_tit_creep}, figure 2 (open circuit).}
 \label{PZT_creep}
\end{figure}
\begin{figure}[H]
\includegraphics[width=0.9\linewidth]{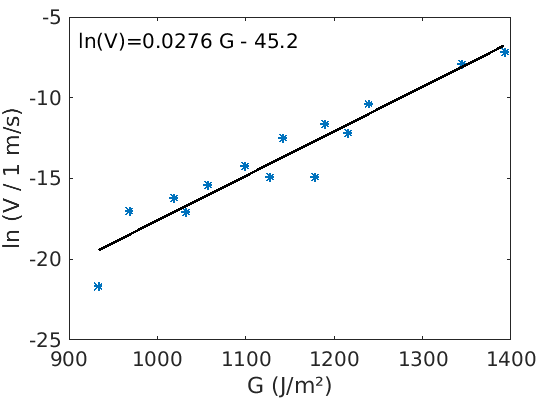}
  \caption{Creep data tetragonal zirconia (TZP) in vacuum, from \citet{TZP_creep}, figure 5.}
 \label{TZP_creep}
\end{figure}
\begin{figure}[H]
\includegraphics[width=0.9\linewidth]{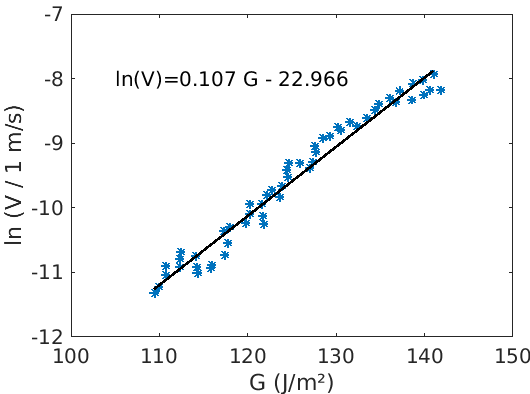}
  \caption{Interfacial creep data in sintered PMMA plates in air, from \citet{Lengline2011}, figure 5.}
 \label{iPMMA_creep}
\end{figure}
\begin{figure}[H]
\includegraphics[width=0.9\linewidth]{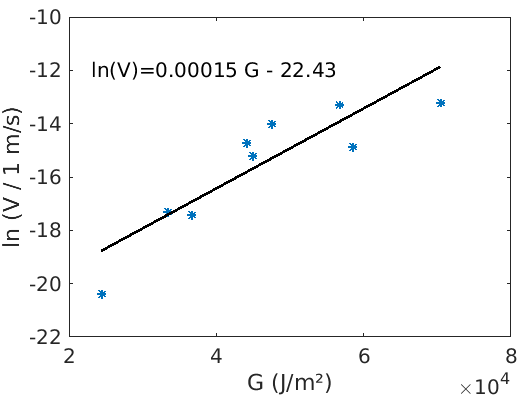}
  \caption{Creep data of high density polyethylene, from \citet{Yoda2001}, figure 4.}
 \label{polyethylene_creep}
\end{figure}
\begin{figure}[H]
\includegraphics[width=0.9\linewidth]{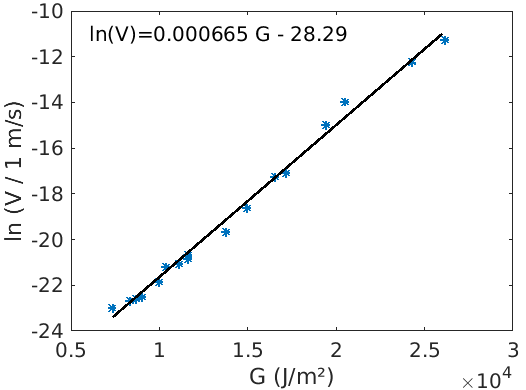}
  \caption{Creep data of aluminium 2650 T6 alloy in vacuum at $175$\,\textdegree\,C, from \citet{alu_creep}, figure 6.}
 \label{alu_creep}
\end{figure}
\begin{figure}[H]
\includegraphics[width=0.9\linewidth]{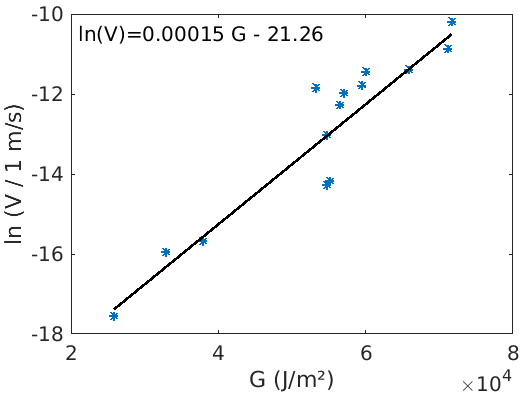}
  \caption{Creep data in Ti-6A1-6V-2Sn titanium alloy in moist air, from \citet{tita_creep}, figure 6a (beta annealed).}
 \label{titanium_creep}
\end{figure}

%------------------------------
\FloatBarrier
%\newpage
\bibliographystyle{unsrtnat}
\bibliography{scr.bib}

% \subsection{Temperature dependant plasticity}

% \begin{figure}[H]
% \includegraphics[width=0.9\linewidth]{granite_creep2.png}
%   \caption{Creep data of Westerly granite in moist air at various ambient temperatures, from \citet{GraniteGabbro_creep}, figure 7.}
%  \label{granite_creep}
% \end{figure}

% \begin{table}[H]
% \begin{tabular}{l|l|l|l|}
% \cline{2-4}
%                                         & \textbf{20 degC} & \textbf{200 degC} & \textbf{300 degC} \\ \hline
% \multicolumn{1}{|l|}{\textbf{$\alpha$ (m$^2$)}} & 3.5E-11          & 5.6E-11           & 8,2E-11           \\ \hline
% \multicolumn{1}{|l|}{\textbf{$G_c$ (J/m$^2$)}}    & 120              & 77                & 42                \\ \hline
% \multicolumn{1}{|l|}{\textbf{U (eV)}}        & 0.9E-10          & 1.5E-10           & 1.8E-10           \\ \hline
% \multicolumn{1}{|l|}{\textbf{l (m)}}        & 8.2E-10          & 3.2E-10           & 1.5E-10           \\ \hline
% \end{tabular}
% \caption{...}\label{table2}
% \end{table}